\documentclass[%
 reprint,
 amsmath,amssymb,
prl
]{revtex4-2}

\usepackage{xcolor}
\usepackage{graphicx}
\usepackage{dcolumn}
\usepackage{bm}
\usepackage{hyperref}

\usepackage{bm}
\usepackage{amsmath}
\usepackage{slashed}
\usepackage{comment}
\usepackage{float}
\usepackage{subfigure}
\newcommand{\braket}[1]{\left<#1\right>}
\begin{document}


\title{Hysteresis of Axionic Charge Density Waves}

\author{Joan Bernabeu}
\email{joan.bernabeu@uam.es}
\affiliation{Departamento de F\'isica de la Materia Condensada, Universidad Aut\'onoma de Madrid, Cantoblanco, E-28049 Madrid, Spain}
\author{Alberto Cortijo}%
\email{alberto.cortijo@csic.es}
\affiliation{Instituto de Ciencia de Materiales de Madrid (ICMM), Consejo Superior de Investigaciones Científicas (CSIC),\\
Sor Juana Inés de la Cruz 3, 28049, Madrid, Spain}%

\date{\today}

\begin{abstract}
Magnetic Catalysis is a known proposal for inducing dynamical axionic gapped phases by means of external magnetic fields from a Weyl or Dirac semimetal phase. At finite Fermi level, the phase transition is of first order type and the magnetic field needs to reach a critical value for the transition to take place. Using the theory of bubble nucleation, we predict the order parameter features a hysteretic behavior as a function of the external magnetic field.  We also analyze the experimental consequences of this hysteretic behavior in several observables like magneto-conductivity, magnetic susceptibility and nonlinear optical coefficients. This hysteretic behavior might serve as a fingerprint of Magnetic Catalysis in Condensed Matter systems.

\end{abstract}

\maketitle

\textit{Introduction.--} Axion insulators are topologically nontrivial quantum states of matter characterized by a magnetoelectric term in their electromagnetic response. For time-reversal invariant topological insulators, this magnetoelectric term is a (topologically quantized) constant coefficient $\theta$. It turns out that, being constant, this term does not modify Maxwell's equations, rendering this term unobservable. To be so, the magnetoelectric coefficient must be space and/or time dependent (axion), $\theta=\theta(t,\bm{r})$, as shown in Ref.\cite{Wilczek87}. There are several ways to observe a nontrivial axionic response from three-dimensional topological insulators (3DTIs): The effect of external magnetic fields, 3DTIs with magnetic surface coatings that gap the surface Dirac state. Another route is to use (time reversal) $\mathcal{T}-$symmetry broken Weyl semimetals, where the axion term is proportional to the separation $\bm{b}$ of the Weyl nodes, $\theta(\bm{r})\propto \bm{b}\cdot\bm{r}$. So far, there are scarce examples of Weyl semimetals with broken $\mathcal{T}-$symmetry. In these cases, $\mathcal{T}-$symmetry is broken due to some sort of antiferromagnetic ordering for Co$_3$Sn$_2$S$_2$ \cite{okamura20} or Mn$_2$Bi$_2$Te$_5$\cite{Zhang_2020}, or the exotic role of large magnetic fluctuations in the case of EuCd$_2$As$_2$ \cite{Ma19}. An appealing consequence of the breakdown of the $\mathcal{T}-$symmetry is that, in these cases, fluctuations around the order parameter might act as dynamical axions, particles hypotethized in the context of High Energy Physics, that couple to electromagnetic fields  through the chiral anomaly-induced term in the effective electromagnetic response of these insulators, if $\theta=\bm{b}\cdot\bm{r}+\delta\theta(t,\bm{r})$, then, the axion mode couples to the electromagnetic fields as $\mathcal{L}_{\textrm{axion}}\sim \delta\theta(t,\bm{r})\cdot\bm{E}\cdot\bm{B}$. In addition to the previously mentioned materials, dynamical axions have been postulated to appear in exotic Charge Density Waves (CDW) from Weyl Semimetal (WSMs) systems \cite{li2010axions,wang2013chiral,roy2015magnetic,roy2017interacting,sehayek2020charge,wieder2020axionic}. Here the CDW sliding mode or phason plays the role of the axion and the Weyl node separation is the CDW wavevector. Two scenarios have been put forward for the axionic phase to appear. The first requires internode interactions to be strong enough so as to lead to the spontaneous symmetry breaking of the global $U(1)$ chiral symmetry \cite{wang2013chiral,roy2017interacting}. The second introduces a strong magnetic field $\bm{B}$ that aids in breaking the symmetry for weaker interactions than in the first scenario \cite{roy2015magnetic}. The latter case is also known as magnetic catalysis (MC) of chiral symmetry breaking \cite{klimenko1991threeI, klimenko1991threeII,gusynin1994catalysis,gusynin1995dimensional,miransky2015quantumfieldtheory} and has been extensively explored in the context of QCD in a strong magnetic field \cite{klevansky1989chiral,suganuma1991on,schramm1992quark,kabat2002qcd,miransky2002magnetic,buividovich2010numerical,delia2010qcd,delia2011chiral,bali2012qcd,bali2012qcd2}. In the context of Condensed Matter physics, the magnetic field is a tuneable parameter, easier to control than fermion coupling strengths, hence MC proves a promising avenue towards observing axionic CDWs.
The substantial difference between these two scenarios for the formation of axionic CDWs is the effective dimensionality in the gap equation. In the latter case, the magnetic field induces the formation of (highly degenerate) $(1+1)$-dimensional Landau level states. The resulting gap equation differs from the one in absence of magnetic field in that the solution implies a non zero value of the chiral symmetry breaking condensate $\Delta=\braket{\bar{\Psi}\Psi}\sim eB \exp(-\frac{1}{G eB})$, for any value of the interaction constant $G$, while at zero magnetic field, this condensate is nonzero for values of $G$ exceeding a critical value 
 $G_c$\cite{wang2013chiral}. 
 For some specific conditions, at zero magnetic field there is the possibility of Fermi surface nesting when a chiral imbalance takes place in the WSM phase\cite{Curtis23}. The presence of a finite chemical potential destroys this nesting condition between Fermi surfaces with opposite chirality, preventing the instability to happen for an homogeneous chiral condensate. Interestingly, it is suggested that a finite-momentum chiral condensate might be allowed in such conditions\cite{Curtis23}.
A similar situation occurs at non-zero magnetic fields. It is already known that temperature or chemical potential restores the chiral symmetry leading to a zero value for the condensate $\Delta$\cite{ebert1999magnetic,Fukushima12}. The salient feature, as we will discuss in the rest of this work, is that, in stark contrast to what happens at finite temperature, at finite chemical potential the effective potential for $\Delta$ develops two minima, one at $\Delta=0$, and the other at finite values of $\Delta$ (in contrast to the $\mu=0$ case where only $\Delta\neq 0$ is a true minimum). The presence of two minima implies that the chiral symmetry breaking phase transition is of first-order type \cite{ebert1999magnetic}. This phase transition has been analyzed in terms of varying the chemical potential at fixed magnetic field. However, in the case of three-dimensional WSMs, it is easier to knob the magnetic field instead of $\mu$. This means that there is now a critical value for the magnetic field $B_0$ below which the chiral symmetry is not broken. This rather trivial conclusion might be behind the MC scenario is difficult to observe experimentally in systems with large $\mu$. Interestingly, that this axionic CDW phase transition is of first order implies a rich structure in the dynamics of the phase transition and allow us to predict phenomenological properties that unambiguously characterize this phase transition.

\textit{Analysis of the Phase Transition.--} From a microscopic point of view, it has been suggested that electron-phonon interactions might be the primary (albeit not unique) sources of electronic correlations in WSMs\cite{zhao21}, either in terms of of a Yukawa-like (screened) coupling between acoustic phonons\cite{Qin20} and electrons, or interactions between electrons and optical phonons\cite{kundu22}.
In any case, for scales smaller than the typical scale of these interactions (screening of the Yukawa interaction or energies below the optical phonon frequencies), it is sufficient with considering a contact interaction. For this reason, we will consider the standard situation of analyzing the Nambu-Jona-Lasinio (NJL) model (a local four-fermion interaction) for Dirac fermions applied to the MC scenario focusing on the infrared properties of the magnetic catalysis of the chiral symmetry. The fermionic Lagrangian of the model at finite chemical potential $\mu$ and magnetic field $\mathbf{B} = B\hat{\mathbf{z}}$ is\cite{ebert1999magnetic, vdovichenko2000magnetic} reads,
\begin{equation}\label{NJL_Lagrangian}
    \mathcal{L}_\textrm{NJL} = \bar{\Psi}(i\gamma^\mu\eta^{\nu}_{\mu}D_{\nu}-\mu\gamma^0)\Psi - \frac{G}{2}\left[\left(\bar{\Psi}\Psi\right)^2+\left(\bar{\Psi}i\gamma^5\Psi\right)^2\right].
\end{equation}
In Eq.(\ref{NJL_Lagrangian}) we have adopted the standard quasi-relativistic notation for gapless fermions in WSMs where $\eta^{\nu}_{\mu}=diag(-1,v_f,v_f,v_f)$, and $v_f$ is the Fermi velocity, so the matrices $\gamma$ follow the conventional Clifford algebra. We
will implicitly take vf = 1, reintroducing it explicitly when convenient. In addition, we define the covariant derivative as $D_\mu \equiv \partial_\mu - ieA_\mu$. We choose to represent the 4-vector potential of a constant magnetic field in the $x_3$ direction in the symmetric gauge, i.e. $A_\mu = \frac{B}{2}(0,-x_2,x_1,0)$. We will assume from this point on that $eB>0$, without loss of generality.

After a Hubbard- Stratonovich transformation that transforms the quartic interaction in Eq.(\ref{NJL_Lagrangian}) into a bilinear interaction \cite{gusynin1995dimensional, ebert1999magnetic} with auxiliary bosonic fields $\Delta$ and $\theta$ such that $\braket{\bar{\Psi}\Psi}\sim\Delta \cos\theta$, and $\braket{\bar{\Psi}i\gamma_5\Psi}\sim\Delta\sin\theta$, one finds that, at finite temperature $T$ and $\mu$, the effective potential for the chiral condensate is the sum of two terms,  $V(\Delta) = V_0(\Delta) + V_{\mu,T}(\Delta)$ where,
\begin{gather}\label{vacuum_contribution}
    V_0(\Delta) = \frac{\Delta^2}{2G} + \frac{eB}{8\pi^2}\int_{\Lambda^{-2}}^\infty \frac{ds}{s^2}e^{-s\Delta^2}\coth(eBs),
\end{gather}
and
\begin{gather}\nonumber
   V_{\mu,T}(\Delta) = - \frac{1}{\beta}\frac{eB}{4\pi^2}\sum_{n=0}^\infty\alpha_n\int_{-\infty}^\infty dp\cdot  \\
   \label{temperature_chemical_potential_contribution}
   \cdot\log\left\{\left[1 + e^{-\beta(\varepsilon_n(p) + \mu) }\right]\left[1 + e^{-\beta(\varepsilon_n(p) - \mu) }\right]\right\}.
\end{gather}
\begin{figure}[ht]
    \begin{tabular}{c}
         \includegraphics[width = 0.5\textwidth]{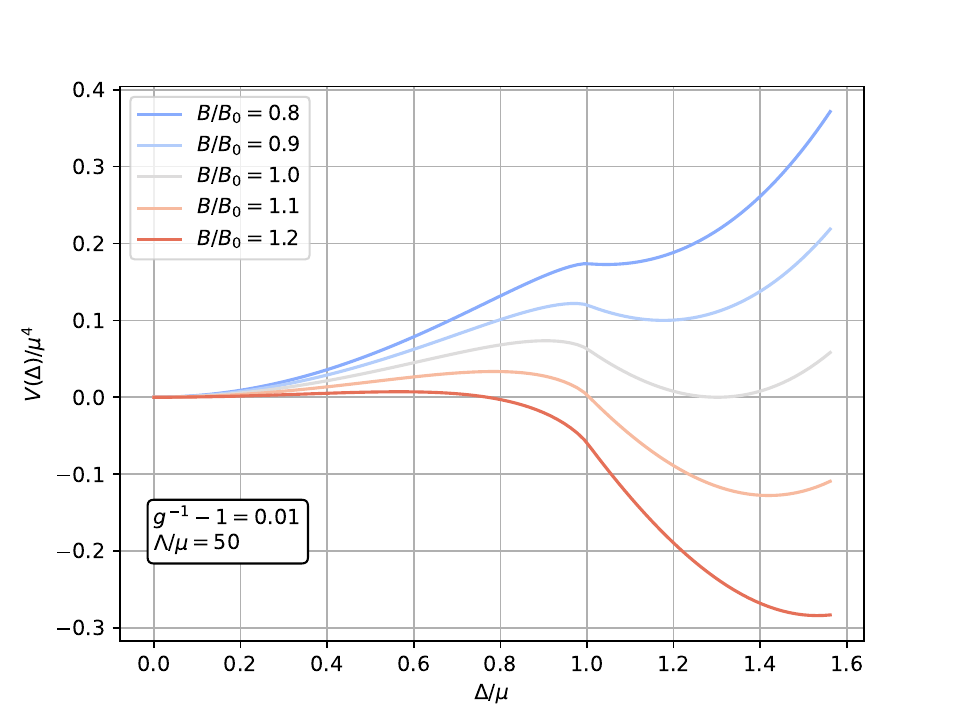} \\
         (a) \\
         \includegraphics[width = 0.5\textwidth]{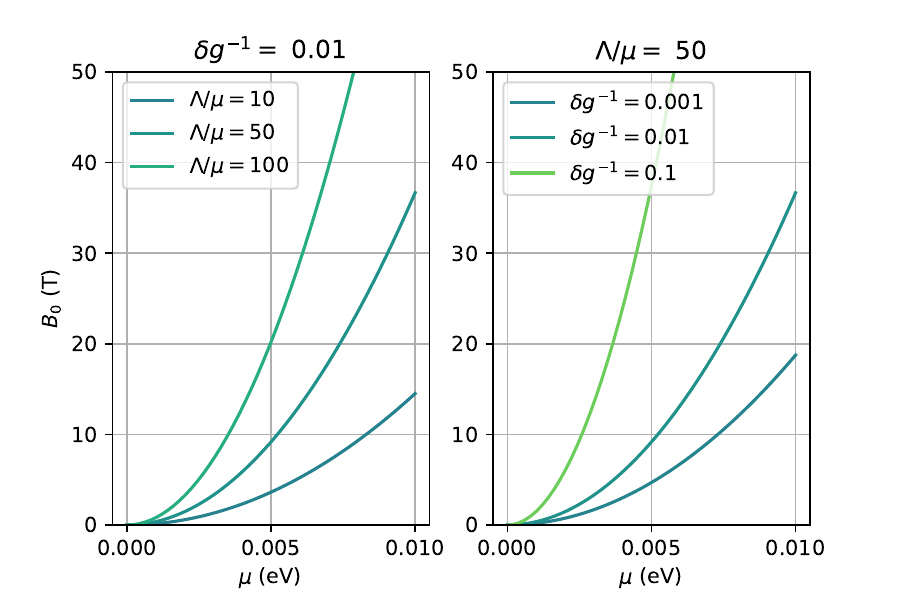} \\
         (b)
    \end{tabular}
    \caption{(a). Plot of $V(\Delta)$ for different magnetic fields, referenced with respect to the degeneracy value $B_0$, where two minima coexist, with $g^{-1} - 1 \equiv \frac{\Lambda^2 G}{4\pi^2} - 1 = 0.01$ and $\Lambda = 50\mu$. (b) Plot of the value of the critical field $B_0$ as a function of the chemical potential for varying values of $\Lambda/\mu$ with fixed $\delta g^{-1} = 0.01$ (left) and varying values of $\delta g^{-1}$ with fixed $\Lambda/\mu = 50$ (right).}
    
    \label{fig:potential}
    
\end{figure}

These expressions come from a standard derivation of the effective action after integrating out the electronic degrees of freedom, and under the assumption that the field $\Delta$ is constant. As usual, the field $\theta$ does not enter in this static approximation as it describes the Nambu-Goldstone mode of the transition.

The term in Eq.(\ref{vacuum_contribution}) is the vacuum contribution that is independent of $T$ and $\mu$, but dependent on the UV-cutoff of the theory, $\Lambda$. $\varepsilon_n(p) \equiv \sqrt{p^2 + 2neB + \Delta^2}$, $\alpha_n \equiv 2 - \delta_{n0}$ are the energies and degeneracies of the $n$ Landau Level, respectively, and $\beta \equiv 1/T$. Within the conventional scenario of magnetic catalysis of chiral symmetry breaking at $T=0, \mu=0$, the minimum of $V(\Delta)$ takes a non-zero value $\Delta_0$ (corresponding to the symmetry broken phase) for any non-zero magnetic field and any $G>0$, hence inducing a gap in the single-particle spectrum. In contrast, in the limit of vanishing magnetic field, a non-zero gap is only reached for couplings above a critical value $G>G_c = \frac{4\pi^2}{\Lambda^2}$. Redefining the coupling constant through the dimensionless parameter $g \equiv \frac{\Lambda^2 G}{4\pi^2}$, this bound conveniently becomes $g> 1$.

There are some qualitative features from the physics of phase transitions that we can draw from Eq.(\ref{vacuum_contribution}) and Eq.(\ref{temperature_chemical_potential_contribution}). Increasing the temperature $T$ from zero but keeping $\mu = 0$, there is a second order (continuous) phase transition at $T \sim \Delta_0$ to the chirally-symmetric phase \cite{Lee98,ebert1999magnetic,miransky2015quantumfieldtheory}. On the other hand, increasing $\mu$ but keeping $T=0$, one finds there is a first-order phase transition for $\mu \sim \Delta_0$. In this regime, the potential (\ref{temperature_chemical_potential_contribution}) takes the form
\begin{eqnarray}\label{chemical_potential_contribution}
   V_{\mu,0}(\Delta) &=& - \frac{eB}{4\pi^2}\sum_{n=0}^\infty\alpha_n\int_{-\infty}^\infty dp \  \Theta\left[\mu - \varepsilon_n(p)\right]\cdot\\
   &\cdot&\left[\mu - \varepsilon_n(p)\right].\nonumber   
\end{eqnarray}
where $\Theta[x]$ is the Heaviside step function. 

For the rest of the present work, we will focus on this quantum regime, $T=0$. The full potential $V(\Delta)$ as a function of the order parameter $\Delta$ at zero temperature but a fixed value of $\mu$ is plotted in Fig.[\ref{fig:potential}.a] for different values of the magnetic field. The main feature is that two minima coexist for a range of magnetic fields. The critical magnetic field $B_0$ is defined as the magnetic field for which the potential minima are degenerate. In addition, we can define two values of the magnetic fields $B_a$ and $B_b$ for which one of the minima disappear: $\frac{d^2V}{d\Delta^2}|_{B=B_{a,b}}=0$ (the symmetric minimum around $\Delta=0$ and the symmetry-breaking minimum at $\Delta\neq 0$, respectively). For magnetic fields $B$ larger than $B_b$ (smaller than $B_a$), the symmetric (symmetry-breaking) minimum  becomes \textit{unstable}. For $B_a < B < B_0$ ($B_b > B > B_0$), the chirality-symmetric (chirality-breaking) minimum is the global minimum, i.e. the \textit{stable} ground state. It is well known that for sufficiently small perturbations around $B_0$ whereby the ground state of the system passes from lying at the global minimum to simply a local one, the system remains in its original, now \textit{metastable}, ground state\cite{Binder87}. In thermal (i.e., temperature-driven) first order phase transitions this phenomenon is known as supercooling (for temperatures $T$ smaller than a critical temperature $T_0$) and superheating (for $T> T_0$). However, the system might transition into the true global minimum well by some external perturbation (or thermal fluctuations in thermal phase transitions), or spontaneously at some values of the magnetic field, $B_1$ and $B_2$, (or any external knob parameter) with $B_a\leq B_1\leq B_0\leq B_2\leq B_b$. In the case where $B_1$ and/or $B_2$ are different from $B_0$, and the transition takes place dynamically by nucleation of bubbles\cite{coleman1977fate1,coleman1977fate2}, the phase transition will show a hysteretic behavior\cite{Agarwal81}. 

All these considerations are generic of first-order phase transitions\cite{Binder87}. In our particular case, the effective potential arises from the pseudo-relativistic interacting electrons described by the model in Eq.(\ref{NJL_Lagrangian}), so it is worth highlighting the formal similarity of this model with the ones studied in the context of cosmological phase transitions \cite{linde1979phase, linde1983decay, csernai1992nucleation, csernai1992dynamics}. Those transitions are often characterized by a symmetry-preserving phase at high temperatures. As the universe expands, it cools down and the temperature is lowered until the symmetry-broken phase becomes energetically favorable at a certain critical temperature $T_0$. If this transition is thermal and of first-order, then for temperatures immediately below $T_0$, the universe will stay in its original but now metastable symmetry-preserving phase. This will occur until bubbles of the symmetry-breaking phase nucleate and expand until they occupy the entire volume of the universe. In these models the nucleation rate $\Gamma$ is given by the high--temperature Arrhenius-like dependence $\Gamma \propto e^{-\Delta F/T}$\cite{langer1969statistical,affleck1981quantum,linde1983decay}. This is in contrast to the quantum scenario considered here, where, as was argued previously, the first-order phase transition is observed when the temperature is the smallest energy scale  $T\ll \mu, \sqrt{eB},\Lambda$. The relevant nucleation rate is then given by the semiclassical expression for $\Gamma$ \cite{coleman1977fate1,coleman1977fate2},
\begin{gather}
    \Gamma \approx \left(\frac{S_b}{2\pi R_b^2}\right)^2 e^{-S_b},
\end{gather}
where $S_b$ is the action for the classical bubble solution to the equations of motion in 4D Euclidean space (that will be specified later), and $R_b$ is the radius of the bubble. Another important difference with respect to cosmological phase transitions is that the transition-driving parameter, the magnetic field $B$, can be lowered or raised in controlled fashion in experiment. Hence, the transition from the chirality-breaking to the chirality-preserving phase might be tuned as the nucleation rate $\Gamma$ will depend on the magnetic field. If the system parameters are such that $(B_1,B_2)$ are different enough from $B_0$, the hysteretic behavior might be observed experimentally. In the following sections we will first estimate the values of $B_1$ and $B_2$ using a modified version of the theory of bubble nucleation and later we will compute several observables where this hysteretic behavior can be measured.


\textit{Calculation of the Nucleation Rate.--} To find the magnetic fields $B_1$ and $B_2$ on either side of $B_0$ for which the material effectively transitions, we calculate the time a critical bubble takes to nucleate \cite{langer1980kinetics, csernai1992dynamics}, defined by $\tau_b^{-1} \sim R_b^3 \Gamma$. We assume that, due to fast expansion of the bubble at speeds approaching $v_f$, the whole sample transitions once a single bubble nucleates. From this perspective, the transition from the symmetric to the broken phase happens at $\tau_b(B_2) < \tau_r$, where $\tau_r$ is some time reference scale much longer than any set in an experiment. Similarly, the magnetic field for which the transition from the broken to the symmetric field occurs $B_1$ is set by $\tau_b(B_1) < \tau_r$. We will consider the conservative $\tau_r = 10^{9}$ years. As will be checked later, $B_1$ and $B_2$ are not too sensitive to smaller, perhaps more sensible, values of $\tau_r$.

To calculate the bubble configuration, we need to take into account kinetic effects beyond the static potential $V(\Delta)$ in Eqs.(\ref{vacuum_contribution},\ref{temperature_chemical_potential_contribution}). This approach has been used for instance to calculate the bubble configurations of first-order chiral phase transition of the early universe \cite{Aoki17}. Keeping the leading terms in the corresponding derivative expansion of the effective action,the kinetic terms for the condensate read
\begin{gather}\label{kinetic_term}
    \mathcal{L}_k = \frac{Z^{-1}(\Delta)}{2}\left[(\partial_0\Delta)^2 - \sum_i v_i^2(\partial_i\Delta)^2\right].
\end{gather}
To recover a canonical $O(4)$-symmetric kinetic term, we introduce a new field $\Phi$ defined by
\begin{equation}
    \frac{d\Phi}{d\Delta} = Z^{-\frac{1}{2}}(\Delta),
\end{equation}
and assume for the sake of simplicity that the velocity of the bosons in the plane perpendicular to the magnetic field $v_\perp = v_1 = v_2$ of the $\Delta$ bosons is equal to the longitudinal velocity, the Fermi velocity $v_f$. This is generally not the case \cite{gusynin1995dimensional, fukushima2013magnetic} as clearly the magnetic field breaks rotation symmetry, but is justified for the cases considered here where the symmetry-breaking ground state $\Delta_0\sim\mu$ is not extremely small compared to $\Lambda$ and the magnetic field is in a non-ultraquantum regime $eB \ll \Lambda^2$ \cite{bernabeu2024chiral}. The field renormalization can be calculated from
\begin{gather}\nonumber
    Z^{-1}(\Delta) = \frac{i}{2}\int\frac{d^4k}{(2\pi)^4}\textrm{tr}\left[\frac{\partial\tilde{\mathcal{D}}}{\partial k_0}\frac{\partial\tilde{\mathcal{D}}}{\partial k_0}\right]= \\
    \label{Z}
    = \frac{eB}{24\pi^2 \Delta^2}\left[ 1 - \left(1 -  \frac{\Delta^2}{\mu^2}\right)^{\frac32}\Theta(\mu - \Delta)\right].
\end{gather}
where $\tilde{\mathcal{D}}$ is the translation-invariant component of the LLL propagator in momentum space \cite{gusynin1995dimensional}, 
\begin{equation}\label{LLLpropagator}
    \tilde{\mathcal{D}}(k) = i\exp\left[-\frac{\mathbf{k}^2_\perp}{eB}\right] \frac{\slashed{k}_\parallel + \Delta}{k_\parallel^2 - \Delta^2}\left(1 - i\gamma^1\gamma^2\right),
\end{equation}
and $k \equiv (k^0,k^1,k^2,k^3)$, $\mathbf{k}_\perp \equiv (k_1, k_2)$ and $k_\parallel \equiv (k_0, k_3)$.
Notice how the $\mu$-dependent term makes the field renormalization finite even in the limit $\Delta \rightarrow 0$, which would otherwise diverge in the $\mu = 0$ case \cite{gusynin1995dimensional}. It is therefore essential for this term to be included, as otherwise if such a divergence were left unabated, then the gapless ground state of $V(\Phi)$ would be found at $\Phi\rightarrow\infty$ and would inhibit any stable bubble configuration with a metastable gapless configuration.

\begin{figure}[ht]
    \includegraphics[width = 0.49\textwidth]{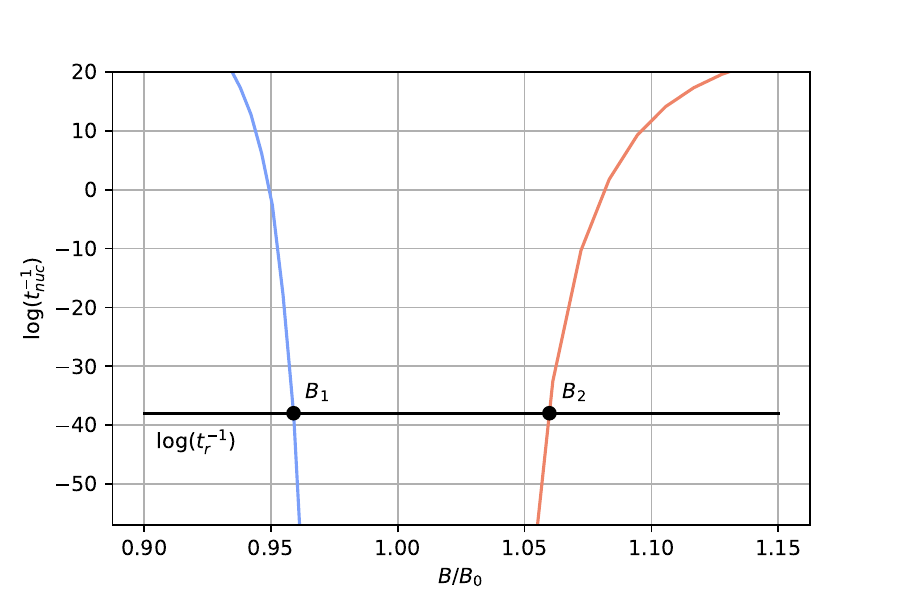}
    \caption{Plot of the logarithm of the inverse nucleation time (in seconds) for different magnetic fields, compared with the reference time $t_r = 10^9$ years, with $g^{-1} - 1 \equiv \frac{\Lambda^2 G}{4\pi^2} - 1 = 0.01$, $\Lambda = 50\mu$ and $\mu = 0.01$ eV.}
    \label{fig:log_times}
\end{figure}

The bubble configurations for different values of the magnetic field can now be obtained solving the equation of motion of the canonical field with respect to the radial coordinate $\rho$, i.e.
\begin{equation}
    \frac{d^2\Phi}{d\rho^2} + \frac{3}{\rho}\frac{d\Phi}{d\rho} =  Z^{\frac{1}{2}}\frac{dV}{d\Delta}.
\end{equation}
imposing the conditions $\lim_{\rho \rightarrow\infty}\Phi = \Phi_\textrm{meta}$, where $\Phi_\textrm{meta}$ denotes the corresponding metastable state, and $d\Phi/d\rho|_{\rho = 0} = 0$, using an undershooting-overshooting algorithm (see e.g. \cite{apreda2002gravitational}). These bubble solutions are then used to calculate the corresponding bubble action
\begin{gather}
    S_b = \int d^4x \left[\frac{1}{2}(\nabla\Phi)^2 + V(\Phi) - V(\Phi_\textrm{meta})\right].
\end{gather}
The resulting nucleation times are plotted in Fig.[\ref{fig:log_times}]. The curves have a steep descent as they approach $B_0$, so much so that lowering $t_r$ down to $10^3$ years, for instance, would not alter the values of $B_1$ and $B_2$ significantly.
Having determined the points at which the phase transition occurs, one can now draw hysteresis curves for gap-dependent quantities. 

\begin{figure*}[ht]
    \centering
    \begin{tabular}{ccc}
        \includegraphics[width = 0.32\textwidth]{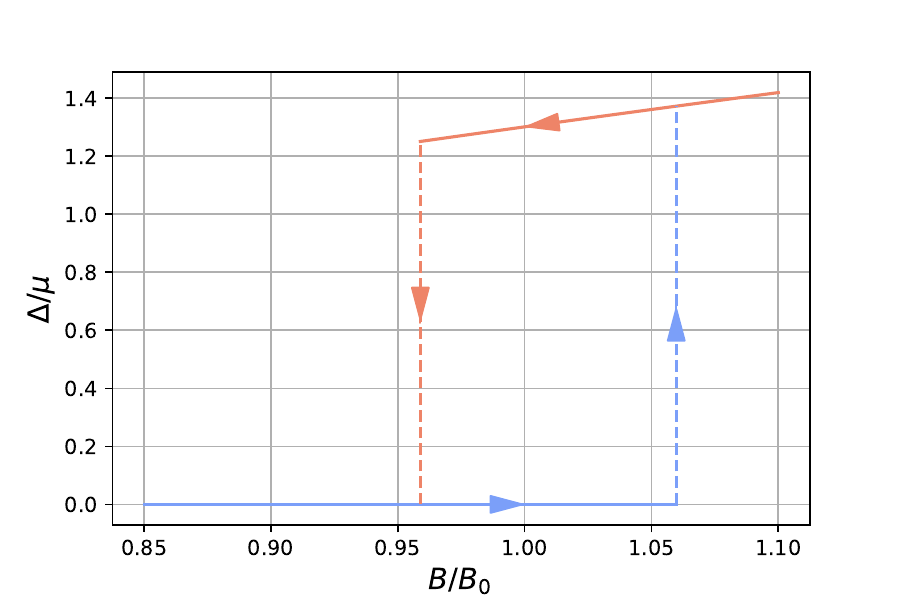} & \includegraphics[width = 0.32\textwidth]{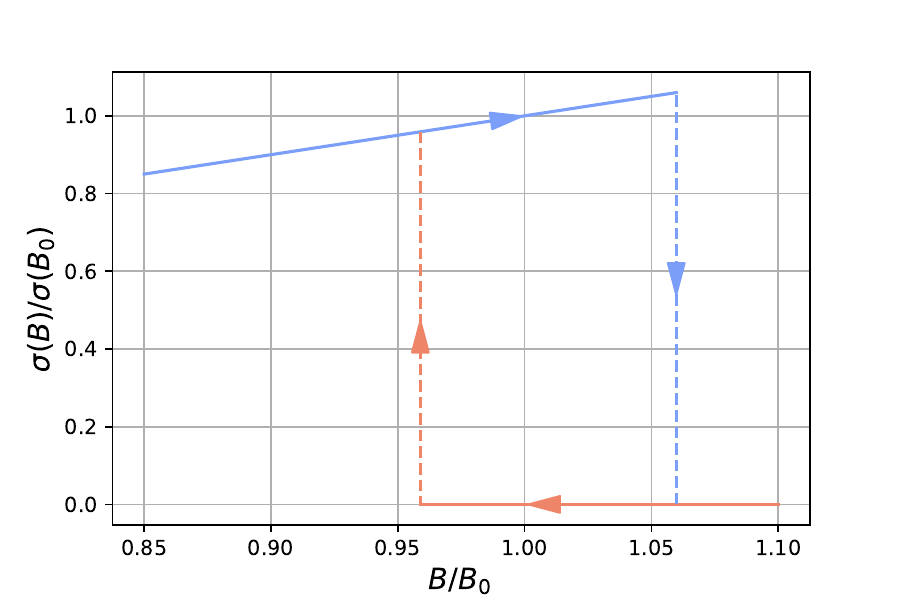} &  \includegraphics[width = 0.32\textwidth]{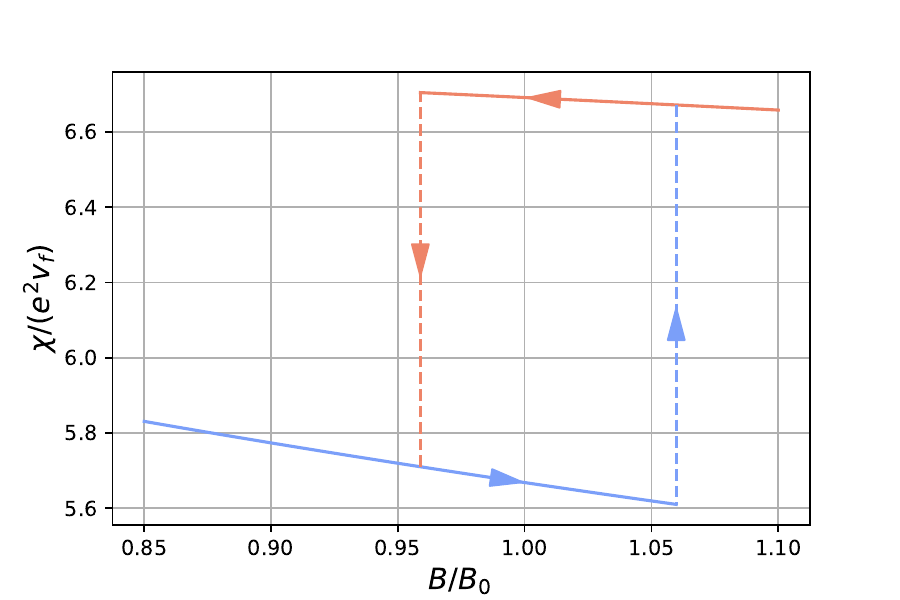}\\
         (a) & (b) & (c) \\
    \end{tabular}
    \begin{tabular}{cc}
         \includegraphics[width = 0.49\textwidth]{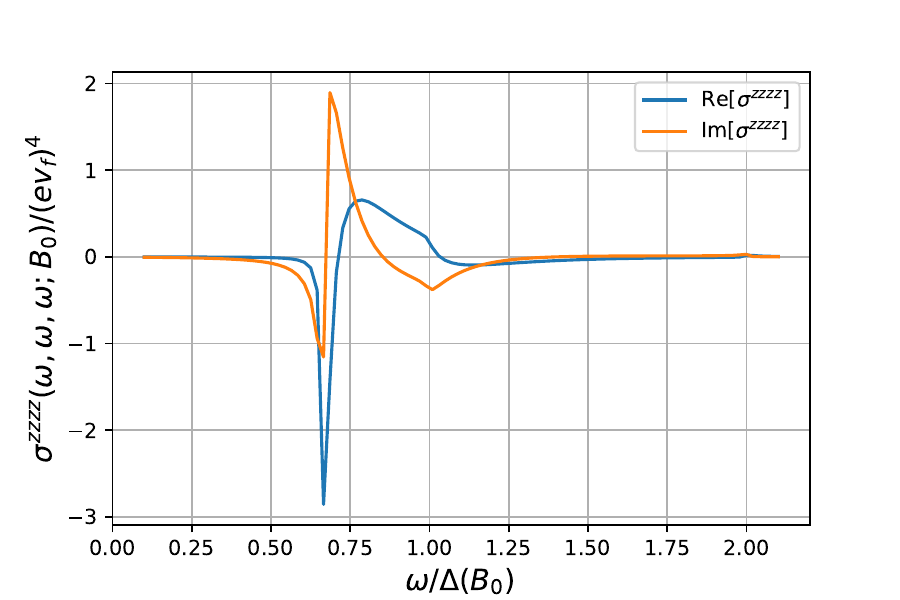} &  \includegraphics[width = 0.49\textwidth]{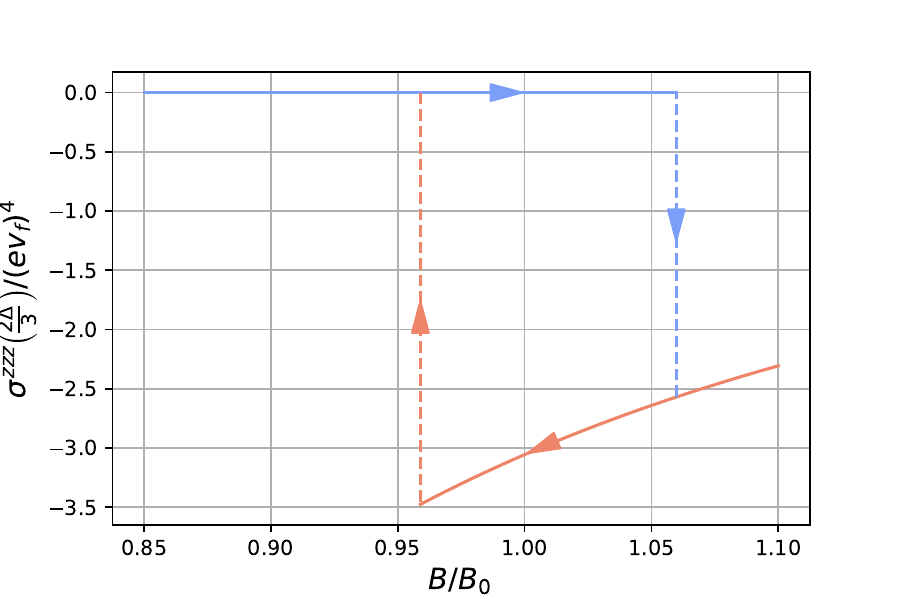}\\
         (d) & (e)
    \end{tabular}

    \caption{Hysteresis diagrams of the gap (a), the magnetoconductivity (b) and the magnetic susceptibility (c). (d) Plot of the third harmonic optical magnetoconductivity component $\sigma^{zzzz}$ at  $B=B_0$. The strong peak corresponds to $\omega = 2\Delta/3$. A broadening of $\eta = 0.01\mu$ (such that $\omega \rightarrow \omega + i \eta$) is employed. (e) Hysteresis diagram of $\sigma^{zzzz}$ evaluated at $3\omega = 2\Delta(B)$. In all plots the parameter values $g^{-1}-1 = 0.01$ and $\Lambda = 50\mu$ have been used.}
    \label{fig:hysteresis_plots}
\end{figure*}
\textit{Observable Consequences.--} In Fig.[\ref{fig:hysteresis_plots}.a-c] we have plotted the hysteresis curves for the gap, the magnetoconductivity \cite{son2013chiral} and the magnetic susceptibility \cite{goswami2011quantum, roy2015magnetic}. Although the model predicts that the system should remain conducting even in the gapped phase because of the surviving Nambu-Goldstone boson, the axion, the associated charge density wave is expected to be pinned if no depinning potential is applied \cite{gooth2019axionic}. This is the case represented in Fig.[\ref{fig:hysteresis_plots}.b].

In addition, we also consider the effects of hysteresis in nonlinear optical response. Since in our model (\ref{NJL_Lagrangian})  the Weyl nodes are located at the same energy, the second-order response is null. Therefore, we consider the 3rd order response, particularly the the component $\sigma^{zzzz}$ at third harmonic generation, where all incident photons have the same frequency $\omega$ \cite{parker2019diagrammatic} and are parallel to the applied magnetic field. The contribution of the LLL, of interest in the magnetic catalysis scenario, is
\begin{gather}\nonumber
    \sigma^{zzzz}(\omega,\omega,\omega) = (ev_f)^4\ \frac{1}{3!}\frac{eB}{4\pi^2}\int_{-\infty}^\infty dp  \ \times \\
    \label{third_harmonic_response}
    \times \frac{2^7\Delta^2(4\varepsilon_0^2(p) - 5\Delta^2 + \omega^2)}{\varepsilon_0(p)(9\omega^6 - 49\varepsilon_0^2(p)\omega^4 +56\varepsilon_0^4(p)\omega^2 - 16\varepsilon_0^6(p))}.
\end{gather}
Note that the denominator can be decomposed into poles around $\omega = \pm 2\Delta/n$ where $n = 1,2,3$. This response function is a sensitive probe to the phase transition as it is zero in the gapless phase. This is fundamentally a consequence of the 1D LLL dynamics and indeed it can be checked that it is nonzero in a Weyl SM or graphene \cite{passos2018nonlinear}  in the absence of a magnetic field. Therefore, as the magnetic field is turned on and increased, one expects to see a gradual decline of $\sigma^{zzzz}$ during the quantum oscillating region until acquiring a minimal value in the quantum limit. However, once the phase transition occurs at $B_2$, it will become finite in the gapped phase, as is represented in Fig.[\ref{fig:hysteresis_plots}.d]. at the critical field $B_0$. The most prevalent feature is the peak at $3\omega = 2\Delta$ in the real part. The hysteresis plot for $\sigma^{zzzz}$ at this frequency is represented in Fig.[\ref{fig:hysteresis_plots}.e].

The dependence of $B_1$, $B_2$ and, fundamentally, of the critical field $B_0$ on the model parameters is explored in Fig.[\ref{fig:potential}.b]. The hysteresis process is accessible for small chemical potentials, as this is when the contribution from the LLL, the key piece in inducing MC, is most relevant. In addition,near-critical couplings $g$ and not so large UV scales $\Lambda$ also favor the process occurring at realistic scales for the magnetic field, as is generally expected in MC  \cite{roy2015magnetic}. It is worth pointing out however that in the ultraquantum limit $\Lambda \gg eB$, no hysteresis should be observed as in that limit the process is independent of the value of the magnetic field modulus \cite{bernabeu2024chiral}.


\textit{Conclusions.--} The present work highlights the fact that a first order phase transition for the magnetic catalysis scenario implies, first, the existence of a chemical potential-dependent critical magnetic field $B_0$ below which the transition cannot take place, and second, a hysteretic behavior in the condensate as a function of the magnetic field, and thus the existence of hysteresis loops in the physical observables that depend on the the condensate, that is only compatible with the effect of many-body interactions. We have shown that a hysteretic process should indeed be observable, as the transition values of the magnetic field, $B_1$ and $B_2$, are well distinguished from one another. The way in which disorder may alter this conclusion remains to be studied in future work. However, it is reasonable to expect it to slow nucleation, and hence widen the hysteretic window, as is the case with 1+0D quantum tunneling \cite{1981_Caldeira_Influence}.

From a experimental perspective in the field of WSMs, the magnetic field-induced axionic phase has been reported in (TaSe$_4$)$_2$I \cite{gooth2019axionic}. Transport measurements in materials in CDW states can however be difficult to interpret due to the Joule heating associated to the voltage to depin the sliding mode \cite{sinchenko2022does}. 
Also, a metal-insulator transition driven by external magnetic fields, compatible with the MC scenario, have been claimed to be observed for the case of ZrTe$_5$ and HfTe$_5$\cite{liu16,tang19,wang20,galeski20}. This point of view has been later disputed in ZrTe$_5$\cite{galeski21}, where no signatures of an electron instability were found, and a conclusive experimental situation for these compounds is still missing. The presence of hysteresis loops in experimental observables might be a conclusive fingerprint of many-body instabilites versus single particle effects.

As a final remark, we note that the pseudo-relativistic theory presented here can be viewed as a condensed matter analogue \cite{barcelo2011analogue} of a cosmological chiral phase transition \cite{csernai1992dynamics,csernai1992nucleation}, with the notable difference that the tuning parameter for the transition is the magnetic field and not the temperature of the expanding universe.

\textit{Acknowledgements.--} J.B. is supported by FPU grant FPU20/01087. A.C. acknowledges financial support from
the Ministerio de Ciencia e Innovación through the grant PID2021-127240NB-I00.
J. B. and A. C. acknowledge discusions with M. A. H. Vozmediano, M. Tolosa-Sime\'on, B. Hawashin, and M.M. Scherer, J.J. Esteve-Paredes, A.J. Uría-Álvarez, M.A. García-Blazquez, and J.J. Palacios.

%

\end{document}